\documentclass[12pt,brazil]{article}
\usepackage[T1]{fontenc}
\usepackage[latin1]{inputenc}
\usepackage[brazilian,english]{babel}
\usepackage{textcomp}
\usepackage{amsmath}
\usepackage{amssymb}
\begin{document}

\title{{\Large Conserved symmetries in noncommutative quantum mechanics}}
\author{V.G. Kupriyanov\thanks{%
e-mail: vladislav.kupriyanov@gmail.com} \\
CMCC, Universidade Federal do ABC, Santo Andr\'{e}, SP, Brazil}
\date{\today                                        }
\maketitle

\begin{abstract}
We consider a problem of the consistent deformation of physical system introducing a new features, but preserving its fundamental properties. In particular, we study how to implement the noncommutativity of space-time without violation of the rotational symmetry in quantum mechanics or the Lorentz symmetry in f{i}eld theory. Since the canonical (Moyal) noncommutativity breaks the above symmetries one should work with more general case of coordinate-dependent noncommutative spaces, when the commutator between coordinates is a function of these coordinates. F{i}rst we describe in general lines how to construct the quantum mechanics on coordinate-dependent noncommutative spaces. Then we consider the particular examples: the Hydrogen atom on rotationally invariant noncommutative space and the Dirac equation on covariant noncommutative space-time.
\end{abstract}

\section{Introduction}

From the point of view of quantum mechanics (QM) the precision of measurement of the position in the space is determined by the wavelength $\lambda$ of the corresponding test particle with the energy $E=h c/\lambda$. Better precision requires higher energies. On the other hand, according to the general relativity, each mass or energy produce the curvature of the space. The Schwarzschild radius corresponding to the energy $E$ is $ r_S(E)={2GE}/{c^4}.$
That is, increasing the quantum mechanical precision of the position measurement one will also increase the Schwarzschild radius of the corresponding test particle. Since one cannot observe nothing inside the black hole, from the condition that $r_S\left({hc}/{\lambda_p}\right)\simeq\lambda_p,$
one f{i}nds the the shortest measurable length $\lambda_p=\sqrt{{G\hbar}/{c^3}}$,
which is also known as a Planck length.

It means that on the very short distances the space-time is nonlocal. This feature can be introduced in the theory supposing that on the Planck scale space-time coordinates $\hat{x}^{\rho},\,\rho=0,1,...,N-1,$ are no longer commutative \cite{Doplicher}. In most simple case one may impose the canonical NC relations,  $ \left[ \hat{x}^{\rho},\hat{x}^{\sigma}\right] =i\theta^{\rho\sigma}\,$ with the constant $\theta^{\rho\sigma}$ being the parameters of noncommutativity.
A problem with the canonical noncommutativity is that it breaks the symmetries of the physical systems, like the rotational symmetry of the Coulomb problem \cite{Chaichian} or the Lorentz symmetry of f{i}eld theoretical models \cite{NCreviews}. The violation of symmetries is easy to understand, the left side of the postulated commutator should transform as a tensor under rotations, while the right side is a constant and does not transform. This fact, leads to the unobservable theoretical predictions imposing the experimental limits on the parameter $\theta$, see e.g. \cite{AGSV} and references therein.

To preserve the symmetries in NC theories we require that the right side of the commutator between coordinates also should be a tensor. For that we admit its dependence on coordinates,
\begin{equation}
\left[ \hat{x}^{\rho},\hat{x}^{\sigma}\right] =i\theta\hat{\omega}^{\rho\sigma}\left(
\hat{x}\right) ,  \label{2}
\end{equation}
where $\hat{\omega}^{\rho\sigma}\left( \hat{x}\right) $ is an operator def{i}ned
from the physical considerations. The consistency condition for (\ref{2}) implies that the symbol of the operator $\hat{\omega}^{\rho\sigma}\left( \hat{x}\right)$ has the form ${\omega}_{q}^{\rho\sigma}\left( {x}\right)=\omega^{\rho\sigma}\left( x\right)+\omega^{\rho\sigma}_{co}\left( x\right)$, where $\omega^{\rho\sigma}\left( x\right)$ is a Poisson bi-vector and $\omega^{\rho\sigma}_{co}\left( x\right) $ stands for non-Poisson
corrections of higher order in $\theta$. These corrections are
expressed in terms of $\omega^{\rho\sigma}\left( x\right) $ and its derivatives
and depend on specif{i}c ordering of the operator $\hat\omega^{\rho\sigma}\left( x\right)$, see \cite{KV} for details and for perturbative construction of the star product corresponding to the algebra (\ref{2}). So, to determine the NC relations (\ref{2})
we should def{i}ne the antisymmetric f{i}eld $%
\omega^{\rho\sigma}\left( x\right) $ and specify the
ordering. In what follows we treat $\omega^{\rho\sigma}\left( x\right)$ as an
external f{i}eld and choose the Weyl ordering.

The QM scale of energies is rather different from the Planck scale. Therefore it is useless to search for the physical effects caused by the noncommutativity in QM. However, it can be used as a toy model to investigate the properties of the proposed model of noncommutativity. The relativistic NCQM can be considered as a f{i}rst step in the construction of NCQFT of general form. NCQM was studied extensively during the last decades, the earlier references can be found in \cite{NCQM}. The model of QM on coordinate-dependent NC spaces of general form (\ref{2}) was considered in \cite{kup14}. Particular examples of NC spaces in f{i}eld theory, such as kappa-Minkowski and fuzzy sphere, can be found in  \cite{Meljanac} and \cite{Vitale} correspondingly.

\section{Quantum mechanics on coordinate-dependent noncommutative spaces}

The general form of nonrelativistic quantum mechanics with NC coordinates was discussed in \cite{kup14,kup15}. The key ingredients of the proposed model are the star product $\star$ which corresponds to the algebra (\ref{2}) and the trace functional $Tr_\star(f)$ on the algebra of this product.  For our purposes we require that the trace should be closed with respect to the star product $\star$,
\begin{equation}
Tr_\star\left( f\star g\right) -Tr_\star\left( fg\right) =0.  \label{trace}
\end{equation}%
The existence of a trace functional satisfying (\ref{trace}), for
a Kontsevich star-product related to any Poisson bi-vector $\omega ^{\rho\sigma}$ was demonstrated in \cite{FS}. The perturbative construction of the trace was proposed in \cite{kup15}.

Let a
function $\mu \left( x\right)\neq 0 $ obeys the equation $\partial _{\rho}\left( \mu \omega ^{\rho\sigma}\right) =0$. The trace functional is de{f}ined as $Tr_\star\left( f\right) =\int d^{N}x\mu \left( x\right) f\left( x\right)$. The problem is that the star product proposed in \cite{KV} does not satisfy the condition (\ref{trace}). To solve it we use a gauge freedom \cite{Kontsevich} in the de{f}nition of a star product. If $\star$ and $\star'$ are two different star products corresponding to the same Poisson bi-vector $\omega^{\rho\sigma}\left( x\right)$, they are related by the gauge transformation
\begin{equation}
D\left( f\star ^{\prime }g\right)=\left( Df\star Dg\right) ,  \label{gauge}
\end{equation}
where $D=1+o(\theta)$ is a gauge operator. In \cite{kup15} it was proposed the perturbative method of the construction of a gauge operator $D$ which relates the given star product $\star'$ with the one $\star$ satisfying the condition (\ref{trace}). So, starting with the star product \cite{KV} and then gauging it with the help of \cite{kup15} we end up with the desirable star product. Up to the second order in $\theta$ it can be written as
\begin{align}
& (f\star g)(x)=f\cdot g+\frac{i\theta }{2}\partial
_{\rho}f\omega ^{\rho \sigma}\partial _{\sigma}g -\frac{\theta ^{2}}{8}\omega ^{\rho \sigma}\omega ^{\alpha \beta}\partial
_{\rho}\partial _{\alpha}f\partial _{\sigma}\partial _{\beta}g \label{star} \\
& -\frac{\theta^2}{12}\omega
^{\rho \sigma}\partial _{\sigma}\omega ^{\alpha \beta}\left( \partial _{\rho}\partial _{\alpha}f\partial
_{\beta}g-\partial _{\alpha}f\partial _{\rho}\partial _{\beta}g\right) -\frac{\theta ^{2}}{24\mu}\partial _{\alpha}\left( \mu \omega ^{\rho \sigma}\partial
_{\sigma}\omega ^{\alpha\beta}\right)\partial _{\rho}f\partial
_{\beta}g +O\left(
\theta ^{3}\right) .  \notag
\end{align}

Having these two ingredients one may proceed with the def{i}nition of NCQM. The Hilbert space is def{i}ned as a space of
complex-valued functions which are square-integrable with a measure $\mu(x)$. The internal product between two states $%
\varphi \left( x\right) $ and $\psi \left( x\right) $ from the Hilbert space
is
\begin{equation}
\left\langle \varphi \right\vert \left. \psi \right\rangle =Tr_\star\left( \varphi
^{\ast }\star \psi  \right).  \label{scalar}
\end{equation}
The action of the coordinate operators $\hat{x}^{\rho}$ on functions $\psi (x)$ from
the Hilbert space can be def{i}ned using the star product (\ref{star}). For any function $%
V\left( x\right) $ we postulate
\begin{equation}
{V}\left( \hat{x}\right) \psi (x)=V(x)\star \psi (x).  \label{15}
\end{equation}
The def{i}nitions (\ref{scalar}) and (\ref{15}) imply that the coordinate
operators are self-adjoint with respect to the introduced internal product (\ref{scalar}), $<\hat{x}^{i}\varphi|\psi>=<\varphi|\hat{x}^{i}\psi>.$
The momentum operators $\hat{p}_{\rho}$ are f{i}xed from the condition
that they also should be self-adjoint with respect to (\ref{scalar}). One may choose it as:
$
\hat{p}_{\rho}=-i\partial _{\rho}-\frac{i}{2}\partial _{\rho}\ln \mu \left( x\right) .
$
The introduced momentum operators commute, $[\hat{p}_{\rho},\hat{p}_{\sigma}]=0$. The commutator between coordinates and momenta is
\begin{equation}
\left[ \hat{x}^{\rho},\hat{p}_{\sigma}\right] =i\delta^\rho_\sigma-\frac{i\theta}{2}\left( \partial_\sigma\omega^{\rho \alpha}\left(\hat{x}%
\right)\hat{p}_{\alpha}+ \frac{i}{2}\partial_\sigma \left(\omega^{\rho \alpha}\partial_\alpha\ln \mu\right) \left( \hat x\right) \right)+O\left( \theta ^{2}\right).  \label{xp}
\end{equation}
So, the complete algebra of commutation relations involving $\hat{x}^{\rho}$ and $\hat{p}_{\sigma}$ is a deformation in $\theta$ of a standard Heisenberg algebra.

As an example we consider the nonrelativistic Coulomb problem. We choose the external antisymmetric f{i}eld in a way to preserve the rotational symmetry of the system,
\begin{equation}
\omega ^{ij}=\varepsilon ^{ijk}x_{k} .  \label{omega}
\end{equation}
The NC coordinates obey the algebra of fuzzy sphere \cite{Fuzzy}, $\left[ \hat{x}^{i},\hat{x}^{j}\right] =i\theta\varepsilon ^{ijk}\hat{x}_{k}.$
Note that the rotationally invariant NC space can be obtained as a foliation of fuzzy spheres \cite{Hammou}.
Due to the choice (\ref{omega}) of the Poisson structure $\omega ^{ij}$, any function $\mu (r^{2})$ obeys the eq. $\partial _{i}\left( \mu \omega ^{ij}\right) =0$ and can be chosen as a measure in the def{i}nition of trace. For simplicity
we set $\mu  =1$. In this case, the momentum operators are just
derivatives $\hat{p}_{i}=-i\partial _{i}$.

Taking into account (\ref{star}) and (\ref{15}) we write the Hamiltonian of the system as:%
\begin{equation}
\hat{H}=\frac{\hat{p}^{2}}{2}- \frac{e^2}{\hat{r}}=-\frac{1}{2}\Delta- \frac{e^2}{{r}}\star= -\frac{1}{2}\Delta- \frac{e^2}{{r}}-\frac{e^{2}\theta ^{2}}{12}\frac{L^{2}}{r^{3}}+O\left( \theta ^{4}\right) ,\label{H}
\end{equation}%
where $r^{2}=x^{2}+y^{2}+z^{2}$ and $L^{2}=L_{x}^{2}+L_{y}^{2}+L_{z}^{2}$ is the orbital
momentum. The conservation of the rotational symmetry of (\ref{H}) implies that the degeneracy of the energy
spectrum over the magnetic quantum number $m$ will be preserved. We use the standard perturbation theory to calculate the leading corrections to
the energy levels,%
\begin{equation}\label{EH}
\Delta E_{n}^{NC}=-\frac{e^{2}\theta ^{2}}{12}\left\langle \psi ^{0}\left\vert \frac{L^{2}}{r^{3}}\right\vert
\psi ^{0}\right\rangle =-\frac{\theta^2 E_n^2}{3 a_0e^2}\frac{n}{l+1/2} ,
\end{equation}%
where $\left\vert \psi ^{0}\right\rangle =R_{nl}(r)Y^m_l(\vartheta,\varphi) $ is
the unperturbed wave function, corresponding to the energy $%
E_{n}=-e^{2}/2a_{0}n^{2}$, $n$ and $l$ are the principal and the azimuthal quantum numbers and $%
a_{0}=1/\alpha $ is the Bohr radius. The contribution of the noncommutativity in the energy spectrum (\ref{EH}) appears as a
correction to the f{i}ne structure of the Hydrogen atom. The corresponding nonlocality is
$
\Delta x\Delta y\geq \frac{\theta ^{2}}{4}\left|m\right| ,
$
where $m=-l,...,l$ and $l=0,1,...,n$.

\section{Noncommutative Dirac equation}

Following the same logic as in the previous section we will introduce the noncommutativity preserving the relativistic invariance in the f{i}eld theory, see \cite{kup17}. The aim of this section is to def{i}ne the relativistic wave equations on coordinate dependent NC space-time which should satisfy the following natural properties: the Lorentz covariance, the continuity equation for the corresponding probability current, and the correct energy-momentum dispersion relation for a free noncommutative relativistic particle, $E^2={\bf P}^2+m^2$.

For simplicity let us consider a $(2+1)$-dimensional noncommutative space-time corresponding to the Poisson structure
$
    \omega ^{\rho \sigma}(x)=\varepsilon^{\rho \sigma\tau}x_\tau.
$
The algebra of coordinates is $ \left[ \hat{x}^{\rho},\hat{x}^{\sigma}\right] =i\theta\varepsilon^{\rho \sigma\tau}\hat x_\tau$.
Since, $\partial_\rho\omega ^{\rho \sigma}(x)=0$, we set $\mu(x)=1$, and $\hat{p}_{\rho}=-i\partial _{\rho}$.
The dynamics of the fermionic f{i}eld $\psi(x)$ interacting with the external vector f{i}eld $A^\rho(x)$ on the coordinate-dependent NC space is described by the following Dirac equation:
\begin{equation}
-\gamma^\rho \hat {p}_\rho\psi+e\gamma^\rho A_\rho\star\psi -m\psi=0,\label{Dirac}
\end{equation}
where $\gamma^\rho$ are the gamma matrices, which in $(2+1)$-dimensions can be represented by the Pauli sigma matrices: $\gamma^0=\sigma_z$, $\gamma^1=i\sigma_x$ and $\gamma^2=i\sigma_y$.
Using the covariance of the Poisson structure $\varepsilon^{\rho \sigma\tau}x_\tau$, one may check that (\ref{Dirac}) is covariant under the corresponding Lorentz transformations, see \cite{kup17}.

For the derivation of conserved probability current for the NC Dirac equation (\ref{Dirac}) we will need the important property of the star product (\ref{star}). The condition (\ref{trace}) implies
\begin{equation}\label{trace2}
 \mu\cdot \left(f\star g\right)=\mu\cdot  f\cdot g+\partial_\rho a^\rho(f,g),
\end{equation}
where
\begin{align}
& a^\rho(f,g)=\frac{i\theta\mu}{4}\omega^{\rho\sigma}( f\partial_\sigma g-\partial_\sigma f g)+\frac{\theta^2\mu}{16}\omega ^{\rho \sigma}\omega ^{\alpha \beta}\left(\partial_\sigma\partial_\alpha f\partial_\beta g-\partial_\alpha f\partial_\sigma\partial_\beta g\right)\label{trace2}\\
&+\frac{\theta^2\mu}{48}\left(\omega^{\beta\sigma}\partial_\sigma\omega^{\alpha\rho}-\omega^{\alpha\sigma}\partial_\sigma\omega^{\rho\beta}\right)\partial_\alpha f\partial_\beta g+O\left( \theta ^{3}\right).\notag
\end{align}
Differentiating the both sides of (\ref{trace2}) and using this equation one more time one f{i}nds:
\begin{equation}
 i\mu\cdot[ (\hat p_\alpha f)\star g]+i\mu\cdot[ f\star (\hat p_\alpha  g)]=\partial_\alpha[ \mu\cdot( f\star g)]+\partial_\rho b^\rho_\alpha (f,g),  \label{p2}
\end{equation}
where,
\begin{equation*}
 b^\rho_\alpha(f,g)=ia^\rho\left(\hat p_\alpha f,g\right)+ia^\rho\left( f,\hat p_\alpha g\right)-\partial_\alpha a^\rho(f,g).
\end{equation*}

Like in the standard relativistic QM we derive the continuity equation from the combination involving the NC Dirac equation (\ref{Dirac}) and its conjugate:
\begin{align}
  -i\mu \bar\psi\star\left(-\gamma^\rho \hat {p}_\rho\psi+e\gamma^\rho A_\rho\star\psi -m\psi \right)\label{id1}\\+ i\mu \left(\hat p_\rho\bar\psi\gamma^\rho+e\bar\psi\star\gamma^\rho A_\rho -m\bar\psi\right)\star\psi&=\notag\\
i\mu\bar\psi\star\gamma^\rho\hat p_\rho\psi+i\mu\hat p_\rho\bar\psi\gamma^\rho\star\psi&=0.\notag
\end{align}
The property (\ref{p2}) implies that the left side of (\ref{id1}) is a total derivative, $\partial_\rho j^{\rho}_\theta,$
where
\begin{equation}
j^{\rho}_\theta=\mu\bar\psi\gamma^\rho\star\psi+b^\rho_\nu(\bar\psi\gamma^\nu,\psi).\label{current}
\end{equation}
Which means that the identity (\ref{id1}) becomes the continuity equation
\begin{equation}
\partial_\rho j^{\rho}_\theta=0,\label{ce}
\end{equation}
for the NC current $j^{\rho}_\theta$. Note that in the commutative limit, $\theta\rightarrow0$ and $\mu=1$, the NC current $j^{\rho}_\theta$ tends to the standard current density, $j^{\rho}_0=i\bar\psi\gamma^\rho\psi$. The NC probability density is
$\varrho_\theta(x)=j^{0}_\theta=\mu\psi^\dagger\star\psi+b^0_\nu(\bar\psi\gamma^\nu,\psi).$ Because of the continuity equation (\ref{ce}) the spacial integral $\int\!\varrho(x)\,d^{2}x$, is a constant in time.

To check the dispersion relation for the free NC relativistic particle we need the def{i}nition of the energy and the momentum of the free NC fermionic f{i}eld $\psi(x)$. Here we use the same logic as in the case of NC current. The identity involving the free NC Dirac equation and its conjugate
\begin{equation}
-i\mu \hat {p}^\nu\bar\psi\star\left(\gamma^\rho \hat {p}_\rho\psi +m\psi \right)+i\mu \left(\hat p_\rho\bar\psi\gamma^\rho -m\bar\psi\right)\star\hat {p}^\nu\psi=0, \label{1d5}
\end{equation}
using the property (\ref{p2}) can be written as
$
\partial_\rho T^{\rho\nu}_\theta=0,
$
where
\begin{eqnarray}
&& T^{\rho\nu}_\theta= \frac{\mu}{2}\left(\bar\psi\gamma^\rho\star\hat {p}^\nu\psi-\hat {p}^\nu\bar\psi\star\gamma^\rho \psi\right) \label{id7}
\\&& -\frac{\mu}{2}\delta^{\rho\nu}\left( \bar\psi\star\gamma^\beta \hat {p}_\beta\psi-\hat p_\beta\bar\psi\gamma^\beta \star\psi-2m \bar\psi\star\psi\right)\notag\\&& +\frac{1}{2}b^\rho_\beta\left(\bar\psi,\gamma^\beta\hat{p}^\nu\psi\right)-\frac{1}{2}b^\rho_\beta\left(\hat{p}^\nu\bar\psi\gamma^\beta,\psi\right) \notag
\\&&
-\frac{1}{2}b^{\rho\nu}\left(\bar\psi,\gamma^\beta\hat{p}_\beta\psi\right)+\frac{1}{2}b^{\rho\nu}\left(\hat{p}_\beta\bar\psi\gamma^\beta,\psi\right)
+mb^{\rho\nu}\left(\bar\psi,\psi\right), \notag
\end{eqnarray}
Note that $T^{\rho\nu}_\theta\rightarrow T^{\rho\nu}_0$ as $\theta\rightarrow0$, where $ T^{\rho\nu}_0=i\bar\psi\gamma^\rho\partial^\nu\psi-i\delta^{\rho\nu}\bar\psi\gamma^\sigma\partial_\sigma\psi+\delta^{\rho\nu} m \bar\psi\psi$. Also $T^{\rho\nu}_\theta$ is conserved, so we identify it with the energy-momentum tensor for the free NC fermionic f{i}eld. The energy and the momentum of the system are $ E=\int\! T^{00}_\theta(x)\,d^{2}x$ and $P_i=\int\! T^{0i}_\theta(x)\,d^{2}x$.

In $(2+1)$-dimensions, when $\mu=1$ and $\hat{p}_{\rho}=-i\partial _{\rho}$, the solution $\psi_{NC}$ of the free NC Dirac equation, $[i\gamma^\rho \partial_\rho -m]\psi_{NC}=0,$ is the same as in the commutative case, except the normalization factor $N$, which should be obtained from the condition, $\int\limits_V\!\varrho_\theta(x)\,d^{2}x=1$. Substituting this solution in the above def{i}nition of the energy and momenta, one obtains the standard energy momentum dispersion relation
\begin{equation}\label{EMrelation}
    E^2={\bf P}^2+m^2+O\left( \theta ^{3}\right).
\end{equation}
We stress that the energy-momentum relation (\ref{EMrelation}) is a consequence of the relativistic invariance of the considered model. If the Lorentz invariance of the external antisymmetric f{i}eld $\omega^{\rho\sigma}\left( x\right)$ will be broken, the energy-momentum relation will change.

In conclusion we note open questions regarding the physical meaning of the proposed model of noncommutativity (\ref{2}). What is the consequence of a combination of noncommutativity with the relativistic covariance. The Lorentz symmetry means that there is no proffered direction in space-time. However, the presence of noncommutativity implies that all coordinates, including $x^0$, are nonlocal. That is the unitarity of the system is broken.

The related problem is the breaking of the translation symmetry with respect to time in the interacting theories. In (\ref{id7}) we have def{i}ned the conserved energy momentum tensor of free NC fermionic f{i}eld. If to introduce in the theory the interaction with an external f{i}eld $A^\rho(x)$, even if $A^\rho(x)$ does not depend explicitly on $x^0$, the translation invariance over $x^0$ will be broken because of the type of noncommutativity. The energy is no longer conserved. The raised questions are applicable not only to the specif{i}c model (\ref{2}) discussed here but to all models of covariant NC spaces, see e.g. \cite{snyder}.

A possible solution of the above problems is the fact that the operator $\hat x^0$, which appears in commutation relations (\ref{2}), does not necessary represent the physical time of the system. As a physical time $\hat\tau$ one may choose the Casimir operator of the algebra (\ref{2}). Since, $\left[ \hat\tau,\hat{x}^{i}\right] =0$, instead of working in coordinates $\hat x^\rho$ we may pass to the new set of coordinates $(\hat\tau, \hat x^i)$ and study the raised questions in these new coordinates.

{\bf Acknowledgement} The paper was supported in part by FAPESP.

%
%

\end{document}